# Formalism to Image the Dynamics of Coherent and Incoherent Phonon with Dark-Field X-ray Microscopy using Kinematic Diffraction Theory

Darshan Chalise[1,2,3], Yifan Wang[1,2,3], Mariano Trigo[2,3], Leora E. Dresselhaus- Marais[1,2,3,*]

[1] Stanford University, Department of Materials Science and Engineering, Stanford 94305, USA
[2] SLAC National Accelerator Laboratory, California 94025, USA
[3] SIMES, Stanford University, Stanford, 94305, California, USA

*Corresponding author: leoradm@stanford.edu

**Abstract**

Dark-field X-ray microscopy (DFXM) is a novel X-ray imaging technique developed at synchrotrons to image along the diffracted beam with a real space resolution of ~100 nm and reciprocal space resolution of ~$10^{-4}$. Recent implementations of DFXM at X-ray free electron lasers (XFELs) have demonstrated DFXM's ability to visualize the real-time evolution of coherent GHz phonons produced by ultrafast laser excitation of metal transducers. Combining this with DFXM's ability to visualize strain fields due to dislocations makes it possible to study the interaction of GHz coherent phonons with the strain fields of dislocations, along with studying the damping of coherent phonons due to interactions with thermal phonons. For quantitative analysis of phonon-dislocation interactions and phonon damping, a formalism is required to relate phonon dynamics to the strains measured by DFXM. In this work, we use kinematic diffraction theory to simulate DFXM images of the specific coherent phonons in diamond that are generated by the ultrafast laser excitation of a metal transducer. We extend this formalism to also describe imaging of incoherent phonons of sufficiently high frequency, which are relevant for thermal transport, offering future opportunities for DFXM to image signals produced by thermal diffuse scattering. For both coherent and incoherent phonons, we discuss the optimal sampling of real space, reciprocal space and time, and the opportunities offered by the advances in DFXM optics.

## 1. Introduction

Dark-field X-ray microscopy (DFXM) is a novel technique that uses an objective lens placed along an X-ray diffracted beam to obtain a full-field, real-space imaging of the diffraction signal from a crystal. DFXM enables X-ray imaging through the depth of the sample with a spatial resolution approaching 150 nm (Simons et al., 2015) with the objective lens, commonly a compound refractive lens (CRL), achieving a magnified image of the sample on a detector (Simons et al., 2017). The CRL also acts an aperture in reciprocal space, and imaging with a reciprocal space resolution better than $10^{-4}$ radians (Poulsen et al., 2021) is possible when using monochromatic and nearly collimated X-rays generated by 4[th] generation X-ray synchrotron and X-ray free electron laser (XFEL) sources. With the excellent resolution in real and reciprocal space and the ability to image bulk crystals in 3D using high-energy X-rays, DFXM can visualize localized strain fields for dislocation structures deep inside bulk materials (Yildirim et al., 2023).

Recent experiments (Holstad et al., 2023 and Irvine et.al 2023) have used DXFM at XFELs to visualize the real-time evolution of strain waves generated by ultrafast laser excitation of a metal transducer. In metal transducers with isotropic thermal expansion (as was the case in the past experiments), the strain wave has a compressive and expansive component along the direction normal to the surface of the metal and propagates away from the surface in the direction of the surface normal vector at the speed of sound. These strain waves are a coherent mode of longitudinal acoustic phonons (Thomsen et al., 1986), which typically have a phonon distribution whose wavelength is peaked in the GHz range. Thus, DFXM images of the strain wave using the femtosecond time resolution available at XFELs allows the visualization of coherent phonon modes and their dynamics in time and space. The ability to visualize the evolution of coherent phonons while also visualizing the strain fields surrounding dislocations provides a direct way to study phonon-dislocation interactions in bulk materials. Additionally, subsurface imaging with high resolution in space and time allows XFEL-DFXM to potentially study the frequency-resolved decay of coherent phonons. This allows quantification of the damping of coherent phonons, which is induced either through the coupling of the strain of the sound wave to thermal phonons (*a.k.a.* Akhiezer damping) or through 3-phonon scattering (Liao et al., 2018). Such measurements of the decay of GHz frequency coherent phonons have important applications in telecommunication (Aigner et al., 2018) and quantum information processing (Bienfait et al., 2019), as it determines the upper bound to Q-value of GHz acoustic resonators.

While GHz coherent acoustic phonons hold important technological relevance for applications in acoustic resonators, incoherent acoustic phonons, especially those with frequencies greater with ≥1 THz frequencies, are the dominant contributors to thermal transport at room temperature in most non-metallic materials (Liao et al., 2018). As such, imaging THz phonons is of importance to thermal transport and thermometry. A significant advantage of lens-based diffraction imaging is that it enables one to spatially resolve even the diffusely scattered light if the lens and the detector are aligned to those beams in the imaging condition (Simons et al, 2017). Therefore, DFXM may be effective to spatially resolve images of thermal diffuse scattering (Xu and Chiang, 2005) from phonons that do not result in a net lattice displacement in any instance of time as opposed to the case of coherent phonons. This diffuse scattering requires more complex equations to describe the scattering signal – beyond the net strain fields used in previous DFXM theory – but is regularly observed in thermal diffuse scattering experiments (Xu and Chiang, 2005).

Holstad et. al (2022) developed a formalism to visualize the propagation of coherent phonons with DFXM when imaging along the main diffraction peak in reciprocal space and along its edges, i.e., at the strong- and weak-beam conditions, respectively. The formalism presents possible geometries for DFXM imaging of coherent phonons and discusses the sensitivities of rocking, rolling and axial strain ($2\theta$) scans. However, to connect that work to applications in acoustics and thermal transport requires a formalism to identify which specific phonon frequencies are described by DFXM images at a specific Bragg condition. Such formalism must quantify the effects of finite real and reciprocal space resolutions and of the sampling rates in time and reciprocal space to inform experimental design to enable studies of frequencies specific to different technologies. In addition, an extension of this to formalism to image thermal diffuse scattering is important for the different applications mentioned above.

In this work, we present a formalism to quantify the measurement of phonon dynamics in bulk single crystals using DFXM within the kinematic theory of X-ray diffraction. We review how strain waves generated by ultrafast laser excitation of a metal transducer can be quantified in a bulk single crystal and use the previous acoustic models to introduce the amplitude and frequency spectra of the coherent phonons transduced into diamond. The resulting DFXM signals are then analyzed based on the goniometer angles and the time resolution required to effectively measure the phonon frequencies relevant to the crystal. We discuss the implications of finite resolution in real and reciprocal space for frequency-resolved measurements of phonon dynamics in single crystalline materials and suggest optimized sampling strategies in space and time. We then extend our framework of DFXM formalism to describe how it may be modified to describe imaging of thermal diffuse scattering from incoherent phonons and discuss the resolution implications for those measurements. The results of this work are intended to guide experimental design and interpretation for visualizing phonon dynamics with DFXM.

## 2. Review of DFXM formalism for kinematic diffraction

We begin this work by reviewing the experimental geometry and formalisms described previously for DFXM modeling of kinematic diffraction, which follows the formalisms introduced in Poulsen, et al. (2021). As shown schematically in Fig. 1a 1D X-ray beam illuminates a single observation plane (shown in blue) that penetrates through the sample (Simons, 2015). The resulting 2D projection of the diffracted light (shown as the orange plane) defines the diffracted beam that propagates along $\vec{k}_d$. The 2D CRL placed along the diffracted beam with sample-CRL and CRL-detector distances that meet the imaging conditions are used to generate the magnified 2D image that describes the spatial extent of the diffraction signal in that observation plane (Simons et al., 2017).

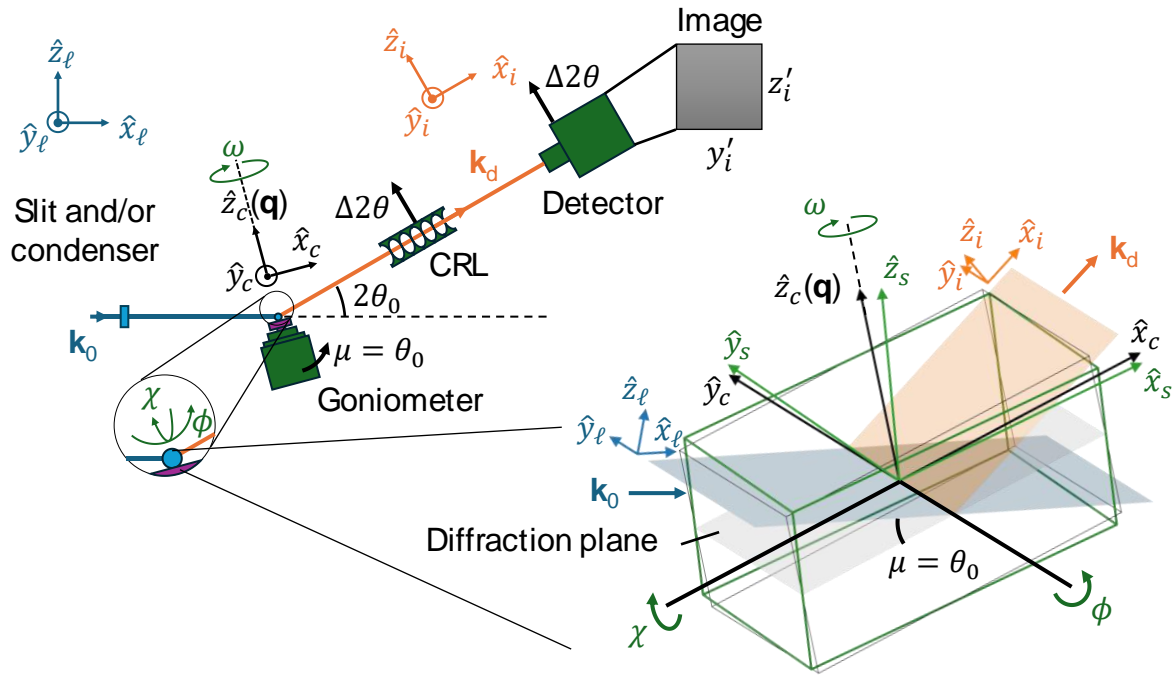

**Fig 1.** DFXM experimental setup (left) and experimental geometry (right). Image adapted from Wang et al. (2024).

Fig. 1 shows the experimental setup and the experimental geometry for DFXM imaging. As the formalism to define the diffraction, goniometer, and crystallography of the system differ based on the scattering signal relevant to the work, it is mathematically convenient to define separate coordinate systems to describe each relevant system. We follow previous formalisms of DFXM (Poulsen et al., 2017 and Poulsen et al., 2021) for these coordinate systems, as described below. The *x*-direction in the lab coordinate system, $x_\ell$, is taken as the direction of the incident X-ray beam. We restrict our description to vertical scattering geometry. The direction along the cross product of the incident and diffracted beam, i.e. the axis of rotation of the diffracted beam with respect to the incident beam is taken as the *y*-direction, $y_\ell$, in the lab coordinate frame while the axis perpendicular to $x_\ell$ and $y_\ell$ is taken as $z_\ell$. It is important to note that in formalisms describing XFEL experiments, $z_\ell$ is used to represent the incident X-ray beam direction (Dresselhaus-Marais, et al., 2023). In this work, we use the synchrotron coordinate system to conform to the previous DFXM theory work, though we note that only a simple coordinate transform is required to use the XFEL system as described fully in Dresselhaus-Marais, et al. (2023).

The direction of the reciprocal lattice vector satisfying the nominal Bragg condition of the selected diffraction peak, $\vec{Q}_{hkl}$, is used to define the *z*-axis of the crystal coordinate system, $z_c$. In the Simplified Geometry described in (Poulsen, et al., 2021), diffraction is defined in the $x_\ell$-$y_\ell$ plane requiring that the reciprocal lattice vector also reside in the same plane. As such, the *y*-axis of the lab and crystal system are collinear, and they are related by a rotation of the nominal Bragg angle, $\theta_0$, about the $y_\ell$-axis, as shown schematically in Fig. 1.

The crystal alignment in DFXM uses a goniometer with 4 axes of rotation: $\mu$ rotation aligns the reciprocal lattice vector into the vertical diffraction geometry, $\omega$ rotates the sample about $z_c$ to rotate the diffraction peak about the Debye-Scherrer ring. The remaining $\phi$ and $\chi$ axes rotate the crystal along mutually perpendicular axes that are nearly orthogonal to the reciprocal lattice vector, with fine rotations for sample scanning. A sample coordinate system is used to relate the lab system to the sample system based on the goniometer angles and their relationship to the normal surface of the crystalline sample. In the simplified geometry as defined in Poulsen et al. (2021) by $\omega = 0$ and $\mu = \theta_0$ when, $\phi = \chi = 0$, the sample and the crystal coordinate systems are collinear with each other at the nominal Bragg condition.

As in the DFXM formalisms previously defined, a vector in the grain system may be converted to the sample system using the coordinate transform matrix, $\boldsymbol{U}^{(g \to s)}$, as $\vec{q}_s = \boldsymbol{U}^{(g \to s)} \vec{q}_g$ The appropriate transformation relating the sample and grain coordinate systems in the geometry we use in our simulations (described in Fig. 2a) is

$$\boldsymbol{U}^{(g \to s)} = \begin{pmatrix} 0 & 0 & -1 \\ 0 & 1 & 0 \\ 1 & 0 & 0 \end{pmatrix}. \quad (1)$$

Samples are scanned by DFXM by collecting a stack of images with the crystal at a series of positions as the rotational angles of the crystal/detector are scanned. Each position of these scans measures a projection of the displacement gradient field onto the reciprocal lattice vector for that orientation. Previous works define a "rocking scan" as a series of images collected as the crystal is rotated in and out of the Bragg condition along the $\phi$ direction, a "rolling scan" as a series of images collected during a $\chi$ rotation in and out of the Bragg condition, and a "strain scan" as a coherent $\theta$-$2\theta$ scan of the crystal, lens, and detector through the axial strain along $\vec{Q}_{hkl}$. A $\theta$-$2\theta$ scan rocking the crystal by an angle of $\Delta\theta$ and the lens and detector by an angle of $2\Delta\theta$ produces the maximum contrast at $\Delta\theta$ values corresponding to axial strain along the diffraction vector. In general, the shear components described by the rocking scans tend to be the most angularly sensitive based on the coherence of the incident beam.

The formalism developed by Poulsen et.al (2021) establishes how DFXM images selected strain fields in the kinematic diffraction limit (i.e. single scattering event per photon). That work implements this as a convolution of a micromechanical model and an instrumental resolution function defined previously in Poulsen et.al (2017) to describe the intensity measured on each pixel of the detector i.e.

$$I(x'_i, y'_i) = \int \Phi^0(r_i) Res_{q_i}(q_i[F^g(r_i); (\varphi, \chi, \omega, \Delta\theta)]) \, dx_i. \quad (2)$$

Here, $Res_{q_i}(q_i[F^g(r_i); (\varphi, \chi, \omega, \Delta\theta)])$ defined the instrument resolution function which also depends on the displacement gradient $F^g$ and the goniometer angles, and $(x'_i, y'_i)$ describe the coordinates of the detector plane in the imaging coordinate system.

In this work, we follow the X-ray scattering formalism described in Poulsen, et al. (2021) but describe an alternate coherent phonon-based model to be used in place of the micromechanical model to describe how DFXM images specific components of the strain fields that define the frequencies relevant to phonon populations. For incoherent phonons, we replace strain fields with specific scattering angles and scattering intensities for specific phonon frequencies resulting in thermal diffuse scattering.

### 3. Imaging of coherent acoustic phonons with DFXM

In this section, we define a material model to image coherent acoustic phonons for the DFXM forward model and demonstrate its utility with our own simulated DFXM images using the formalism described in Section 2. We discuss the advantages of this approach and the considerations for real- vs reciprocal-space resolution in designing such experiments. For our formalisms, the material models are defined in the grain coordinate system to simplify the conversions between the diffraction and material models.

### 3.1 Review of coherent acoustic phonons production in metal transducers and their transduction in the sample

As the phonon experiments to date with DFXM have focused on laser-based excitation of acoustic waves in a transducer, we begin our formalism with a review of the models used to define the phonon modes in the photothermal generation of acoustic waves described previously. We begin with a discussion of the expected characteristics for coherent phonons produced from ultrafast excitation of a metal transducer. In the sections that follow, we will use the models presented in this section to describe the strain fields comprising of coherent acoustic phonons and use this as a new approach to the micromechanical models used to describe DFXM signal in the forward model.

When a femtosecond laser pulse is absorbed at the free surface of a metallic layer, the gradient in thermal energy deposited into the metallic lattice due to the finite penetration depth of the laser initiates a strain wave that propagates along a vector normal to the surface at the longitudinal velocity of sound (Thomsen et al., 1986). The resulting strain wave can be described as a quasi-1-dimensional wave that propagates in the direction normal to the excitation surface (defined as the *x*-direction in our grain system). The propagation direction is a result of the comparison between the spatial extent of the laser spot (~µm) and the penetration depth of the visible laser into the metal (<100 nm). The laser-induced propagating strain wave represents a coherent superposition of longitudinal acoustic phonons that generate a non-zero net displacement of the lattice at a particular depth and time (Lindenberg et al., 2000). For metals with ultrafast electron-phonon coupling, the time domain profile of the longitudinal strain is described by the commonly used Thomsen model (Thomsen et al., 1986)

$$\varepsilon_{xx}(x,t) = (1-R)\frac{Q\beta}{A\xi C}\frac{\rho v^2}{3B}[e^{-\frac{x}{\xi}}\left(1-\frac{1}{2}e^{-\frac{vt}{\xi}}\right) - \frac{1}{2}e^{-\frac{|x-vt|}{\xi}} sgn\,(x-vt)]. \quad (3)$$

Here, $\varepsilon_{xx}(x,t)$ represents the tensile strain along the $x$ axis at depth $x$ below the free surface, $R$ is the coefficient of reflectivity of the metal for the selected laser wavelength, $Q$ is the laser energy per pulse, $\beta$ is the coefficient of thermal expansion, $A$ is area of the laser spot on the sample determined by the $1/e$ radius of a Gaussian beam, $C$ is the mass specific heat capacity of the metal, $B$ is the bulk modulus, $\rho$ is the density and $\xi$ is the optical penetration depth of the laser on the metal.

The characteristic phonon frequency predicted by the Thomsen model results in a distribution that peaks at a characteristic frequency of $f \sim \frac{\xi}{v}$. This has been experimentally observed in metals like nickel (Ni) which have strong electron-phonon coupling (Saito and Wright, 2003). In metals like Ni and Co (Ruello and Gusev, 2015), the characteristic frequency of coherent phonons is ~ 100 GHz as the laser penetration depth is ~10 nm.

For metals like gold and aluminum, where electron-phonon coupling is weaker, the electrons from the laser excitation diffuse to distances longer than the optical penetration depth before they transfer their energy as heat in the lattice. For these metals, the characteristic phonon wavelength is given by $l_c = \sqrt{\frac{k}{g}}$ for an electronic thermal conductivity of $k$ and the electron phonon coupling constant $g$. For metals with $l_c = \sqrt{\frac{k}{g}} \gg \xi$, the strain profile taking into account the electron-phonon coupling constant is described by Wright and Gusev (Wright & Gusev, 1995; Wright & Gusev, 1996),

$$\varepsilon_{xx}(x,t) = (1-R)\frac{Q\beta}{AC}\sqrt{\frac{g}{k}}\left[e^{-\frac{|x-vt|}{\sqrt{k/g}}} sgn\,(x-vt)\right]. \quad (4)$$

The Gusev-Wright model accounts for finite diffusion of hot electrons, and for gold it predicts a strain profile that peaks for phonons at frequency ~10 GHz.

Based on the Nyquist-Shannon sampling theorem (Shannon, 1949), experiments studying coherent phonons produced in ultrafast electron-phonon coupling metals like Ni require sampling in time domain in steps smaller than 5 ps to reconstruct phonons with frequencies ~100 GHz. By contrast, experiments on metals with low electron-phonon coupling like Au require only sampling steps of ~50 ps to achieve the characteristic oscillations of the relevant phonons.

When designing experiments tailored to study specific phonon frequencies, one also must be aware that the expected phonon spectra predicted by Gusev's or Thomson's model is only true if the thickness of the metal transducer is much greater than $\xi$ and $\sqrt{\frac{k}{g}}$. If the thickness of the transducer is smaller than $\sqrt{\frac{k}{g}}$, the characteristic phonon wavelength is instead determined by the thickness of the transducer (Daly et al., 2009) for non-metallic samples, as the electron diffusion is limited to the thickness of the metal transducer.

For DFXM experiments, the laser-transducer interaction generates the phonon frequencies that arrive to the interface between the sample and the metal. However, imaging experiments are typically performed in an adjacent crystalline material – relying on transduction of the generated photothermal wave into the crystal of interest. To describe the conductance of phonons across the metal-crystal interface, one must account for accurate transduction to describe the strain profile in the sample. The roughness of the material interface is perhaps the most important feature in defining the most appropriate model to describe the transduced phonon frequencies. For phonon wavelengths much larger than the roughness of the interface, typical for a well-polished sample, the acoustic mismatch model provides a reliable way to calculate the phonon conductance across the interface (Cahill et al., 2003). For all phonon frequencies, the acoustic mismatch model predicts a phonon conductance across the metal layer 1 and sample layer 2 by: $T = (Z_1 - Z_2)/(Z_1 + Z_2)$, where $Z = v\rho$ is the acoustic impedance of a layer with a speed of sound $v$ and a density $\rho$. Since in this case, we consider a single-crystal diamond, which are typically well-polished, we assume the acoustic impedance model to be effective to describe the transduction. For phonon wavelengths that are similar to or smaller than the roughness of the interface, more complex models must be used that are beyond the scope of this work (Cahill et al., 2003).

Finally, when calculating the strain profile across an interface and in the sample of interest, the change in acoustic velocity must also be included. To account for the change in acoustic velocity and amplitude across the interface and maintain the frequency distribution of the phonons, we introduce a modified strain-wave equation for sample layer 2, given by

$$\varepsilon_2(x,t) = (1-R)\frac{TQ\beta}{AC}\sqrt{\frac{g}{k}}\left[e^{-\frac{v_1|z-v_2 t|}{v_2\sqrt{k/g}}} sgn(x-v_2 t)\right]. \quad (5)$$

From the analytical expression in Equation (5), it becomes evident that the strain wave generated by the laser excitation has a bipolar nature, with expansive (positive) component beyond the spatial position $x = v_2 t$ and a compressive (negative) component before the spatial position $x = v_2 t$ after time $t$ of the strain wave launching into the sample.

### 3.2 Integrating imaging of coherent phonons into DFXM formalism

In this section, we adapt the DFXM simulation formalism derived by Poulsen et al. (2021) to our coherent phonon generation scheme. Our study will focus on DFXM image simulations of the axial strain along the $\vec{Q}_{hkl}$ diffraction vector, as described in Section 3.1, to image an acoustic wave in diamond, produced in a preceding aluminum transducer. As the bipolar strain wave propagates normal to the surface on which the transducer is deposited, the grain system offers a convenient coordinate system to define the wave generation. Even if the laser beam is incident at an oblique angle to the sample, the sample's surface heating still exhibits a shallow cylindrical geometry (i.e. "pancake-like") that causes the strain wave to still propagate normal to the sample surface (Cahill et al., 2003). Our grain coordinate system in this case is defined such that the unit vectors $\hat{x}_g, \hat{y}_g, \hat{z}_g$ point along the crystallographic directions of diamond i.e. (1 0 0), (0 1 0) and (0 0 1) respectively.

While previous experiments have used X-ray and laser beams that propagate along nearly the same incident direction, in this work we select the alternate geometry shown in Fig. 2. To simplify the conversion between the measured distortion along the $\Delta\theta$-$2\Delta\theta$ scan in DFXM and the axial strain ($\varepsilon_{xx}$) that is produced by the bipolar strain wave in the sample, we assume the observation plane to be nearly perpendicular to the direction of the propagating strain wave (Fig. 2a). We simulate the imaging using a diffraction from the (4 0 0) peak in diamond, and consequently, the $\hat{z}_c$ vector along the (1 0 0) direction of the undeformed crystal. The bipolar strain wave is launched along the (1 0 0) direction. We assume $\omega = 0$ such that $\phi$ scan represents a rocking scan and a $\theta$-$2\theta$ scan represents an axial strain scan to be measured along the (4 0 0) plane of diamond.

Fig. 2b includes simulated DFXM images that visualize strain waves in diamond that are produced in an aluminum transducer at an X-ray probe time delay of 20 ps and 50 ps with respect to the time when the strain wave reaches into the sample, respectively. Each time delay includes three images simulated for specific positions along the rocking curve, at $\phi = 0$, +0.001 and -0.001 radians, from left to right.

For the reconstruction of the full profile of the time dynamics of the axial strain along the profile simulated (4 0 0) plane, the measurement in $\phi$ should include the range where the maximum and minimum $\phi$ values correspond to the maximum and minimum expected strain in the specific experimental geometry. In the specific geometry of Fig. 2b, utilizing $\phi$ rotations of the sample and $2\theta$ rotations/translations of the lens/detector along the diffraction arc to measure the axial strain, the average strain value for a particular pixel is given by

$$\varepsilon_{xx} = \frac{\phi_{max}}{\theta_0} \quad (6).$$

where $\phi_{max}$ corresponds to the weighted $\phi$ value giving maximum contrast for the selected pixel, and $\theta_0$ corresponding to the nominal Bragg angle.

To generalize the simulation, one needs to ensure the normalized undeformed scattering vector $\vec{q}_{hkl}$ is defined using the basis vectors defining the grain system. One also needs to select the appropriate transformation matrix, $U^{(g \to s)}$, to relate the grain and sample systems that define the orientation of the sample for specific goniometer angles (Poulsen et al., 2021). In our simulation, the $\phi$ motor of the goniometer fully captures the $\varepsilon_{xx}$ strain values of the sample due to the projected along the (400)) diffraction peak. Even in a general geometry for which the launched strain wave is not purely axial along the imaged peak, a DFXM scan of $\theta$-$2\theta$ is sufficient to sample the frequency spectrum of the phonons as long as the strain has a non-zero component along the axial direction. We perform our simulations with the parameters for reciprocal space resolution identical to one used in Poulsen et al. (2021) with incoming beam divergence of $5.3 \times 10^{-4}$ radians in the vertical direction and $1 \times 10^{-5}$ radians in the horizontal direction. The relative energy bandwidth used is $6 \times 10^{-5}$, the physical aperture of the CRL is $4.5 \times 10^{-4}$ m and the sample to objective distance is 0.274 m. We adjust the scattering angle ($\theta$) in simulating the reciprocal space resolution to 31.4 ° to reflect the nominal Bragg condition when using 10 keV x-rays to image the (4 0 0) peak in diamond.

We then express the displacement gradient tensor $F^g$ for our analysis to account for the time dependent axial strain as describe by Equation (2) with $F^g = \begin{pmatrix} 1 + \varepsilon_{xx} & 0 & 0 \\ 0 & 1 & 0 \\ 0 & 0 & 1 \end{pmatrix}$ with $\varepsilon_{xx}$ described by Equation (2). We use the transformation matrix $U^{(g \to s)}$ as described by Equation (1) to transform from grain to sample coordinate system.

For capturing the time dynamics, we perform the simulation with time step of 1 ps and a time range of 0.6 ns which allows reconstruction of phonon frequencies from 2 GHz to 500 GHz. For each time step, we simulate DFXM images for $\phi$ values ranging from -0.012 radians to + 0.012 radians in 200 steps.

All our simulations are performed assuming a laser excitation of 800 nm, with per pulse energy of 22 µJ and a laser spot size diameter of 180 µm and an aluminum transducer, corresponding to the experimental parameters described in our previous DFXM experiment (Irvine et al., 2023). We ignore the strain wave dissipation in the transducer, assuming the transducer is thin (of the order of 100 nm) and our strain model described in Equation (2) does not take into account the finite pulse duration of the laser (usually of the order of femtoseconds).

In the subsections that follow, we use these imaging results to discuss the implications of finite resolutions DFXM's ability to resolve the phonon spectra of the acoustic waves.

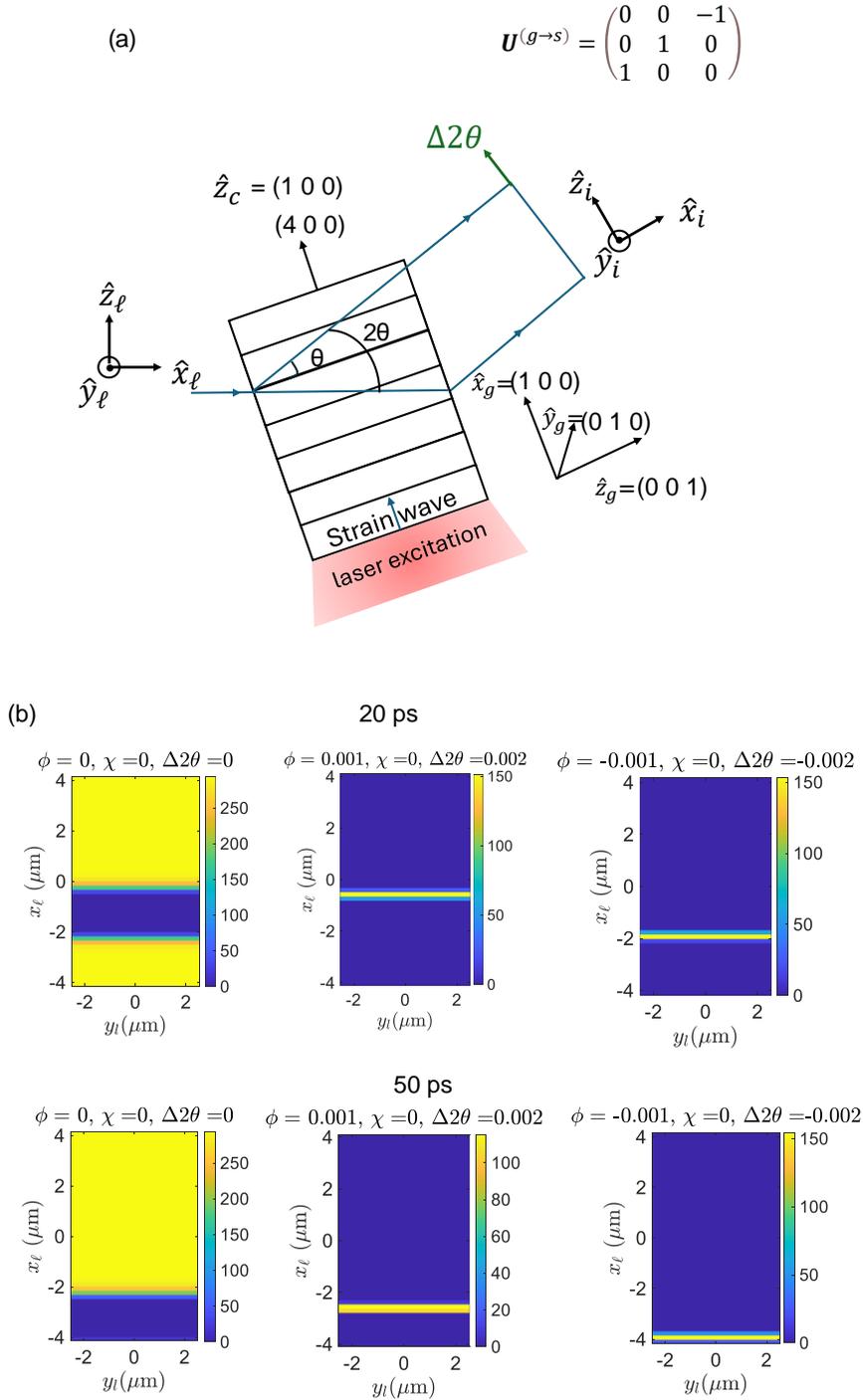

**Fig 2.** Visualization of coherent phonon propagation with DFXM. (a) shows the geometry used in our simulations for the visualization. The crystal surface directions used for the simulation are along the three crystallographic axes. The diffraction is collected from a (4 0 0) peak while the strain wave is launched along the (1 0 0) direction. (b) shows the visualization of the strain wave propagation in diamond XFEL delay times of 20 ps and 50 ps at strong beam and weak beam conditions. The simulated strain wave is produced in an aluminum transducer with an 800 nm laser excitation. The intensities on the right of every image in (b) is arbitrary units scaling with expected intensity. The color bar for image scales with the maximum and minimum intensities for every individual image.

### 3.3 Implications of finite real and reciprocal space resolution, illumination X-ray beam width and sampling

*Finite real space resolution*: Fig. 3a shows the profile of the actual strain wave in diamond produced in an Al transducer using an 800 nm ultrafast laser. The average strain profiles in a space of 250 nm and the strain reconstructed from weighted $\phi$ values from a single pixel in a θ-2θ axial strain scan simulation are also included for comparison. Fig. 3b shows the frequency decomposition of the strain profile. Higher frequency components of the phonon spectra, which have a spatial wavelength smaller than the pixel size get smeared out due to the finite spatial resolution, i.e. the integration over the sample signal that defines the finite pixel size in the object plane for a DFXM measurement. From the sampling theorem (Shannon, 1949), the limit to successful frequency reconstruction of coherent acoustic phonons due to finite real space resolution is $f \leq \frac{v_s}{2x_{gp}}$ for a speed of sound in the material $v_s$ and a pixel size $x_{gp}$ in the grain coordinate system. The results of this under sampling are shown by the plots in Fig. 3a-b, which illustrate that the strain field (orange) is under sampled by the pixel averaging (blue, green). For the DFXM simulated reconstruction of the frequency profile in Fig. 3b, it becomes evident that the *measured* peak phonon population is shifted to lower frequencies due to the 250 nm pixel averaging in the grain system, corresponds to a maximum observable frequency of ~30 GHz.

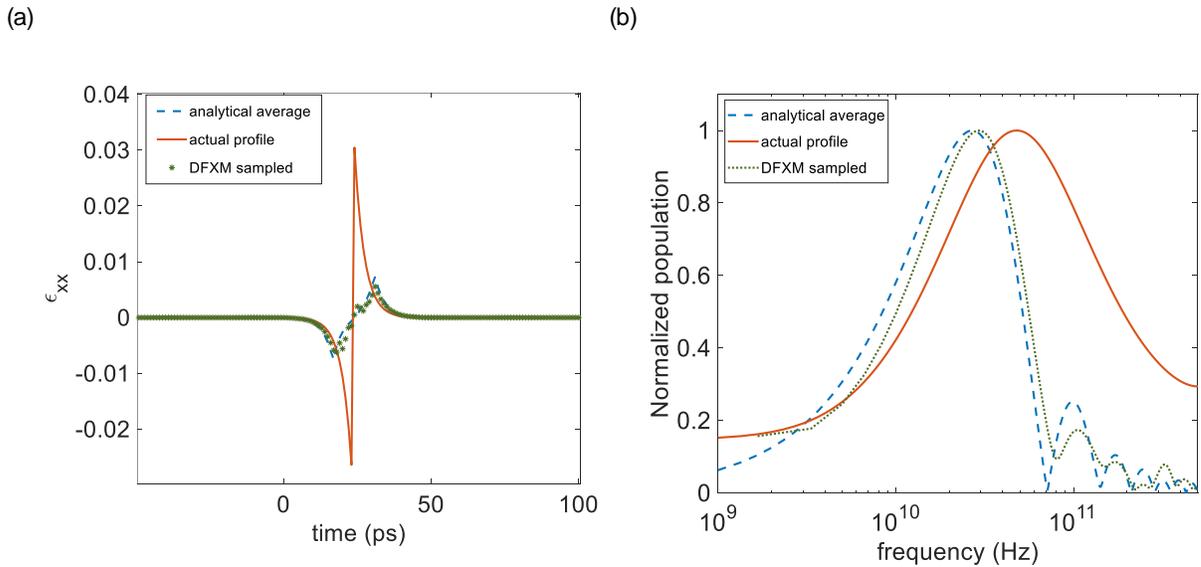

**Fig 3.** Comparison of coherent phonon profile produced in Al, imaged in diamond. (a) shows the time domain of the strain. The expected profile for an infinitesimal point from the Gusev-Wright model is shown as the orange solid line, the expected average profile in a 250 nm pixel is shown as the dashed blue line and the strain corresponding to the weighted $\phi$ value for maximum intensity in the DFXM simulation with 250 nm resolution in the grain system is shown as green stars. (b) shows the reconstructed frequency profile of the phonons comparing the actual frequency spectrum from the Gusev-Wright model (red solid line), the analytical average with a 250 nm resolution (blue dashed line) and the spectrum reconstructed from the DFXM simulation with a resolution of 250 nm in the grain system (blue dashed line).

Therefore, in diamond, studying coherent longitudinal phonons produced in low electron-phonon coupling transducers (for e.g. gold which peaks at ~10 GHz for 800 nm) necessitates a spatial resolution (object plane pixel size) of ~500 nm while studying coherent phonons produced in high electron-phonon coupling transducers (e.g. Ni which peaks at ~100 GHz for 800 nm) necessities a spatial resolution of ~50 nm. These requirements are doubled in Si due to the lower value of the speed of sound.

As the theoretical limit to the spatial resolution in DFXM is given by $\Delta x \approx \frac{\lambda}{3\sigma_a}$ for an X-ray wavelength $\lambda$ and a CRL aperture of $\sigma_a$, one could possibly improve the limit of frequency reconstruction by either performing the experiment at higher X-ray energies or with focusing optics that offer a larger aperture. As discussed in the following section, however, increasing the lens aperture compromises the reciprocal space resolution, which is detrimental to the distinguishability of low frequency phonons in the spectrum. Therefore, ideal experiments to study phonon dynamics require X-ray lenses with high magnification at small X-ray wavelengths without compromising the reciprocal space resolution.

*Finite reciprocal space resolution*: In Fig. 4, we demonstrate an analytical comparison of the time and frequency domain strain profiles in diamond produced with the ultrafast excitation of aluminum transducer with pump laser energy of 22 µJ and 2.2 µJ in a spot of 180 µm diameter, respectively, to investigate the effect of finite reciprocal space resolution. To decouple the effect of real space resolution in frequency reconstruction we perform all these analysis with the assumption of infinitesimal real space resolution. The time and frequency domain profiles with infinitesimal real and reciprocal space resolution for 22 µJ and 2.2 µJ laser energies are shown as orange lines in Fig. 4 (a-d).

Using the reciprocal space resolution function using the parameters described in Section 3.2, the strong beam condition i.e. high intensity in DFXM image when no strain is present, occurs in the range of $\phi = -0.0004$ to $\phi = +0.0004$. From Equation (4), this implies that axial strain smaller than $\frac{0.0004}{\theta_0} = 1.6 \times 10^{-3}$ cannot be distinguished in the DFXM simulation. Therefore, in Fig. 4, we also include profiles (shown as blue lines) where only the strain values above $1.6 \times 10^{-3}$ are sampled.

In the case of a high laser fluence, the limited reciprocal space resolution does not affect the reconstructed profile significantly. On the other hand, the profile with less laser fluence (2.2 µJ) is significantly affected.

As seen in Fig. 4, the ability to distinguish lower frequency components increases linearly with the increase in fluence of the excitation laser - the magnitude of the slow rising components of strain increases linearly with the increase in power per unit area. Therefore, to resolve the lower frequency components of the coherent phonon spectrum, an ideal experiment involves using excitation fluences just below the damage threshold of the transducer (assuming the strain response is linear with the laser excitation).

Additionally, the improvement in reciprocal space resolution, through decreasing in energy bandwidth and incoming beam divergence can improve the ability to resolve low frequency phonons.

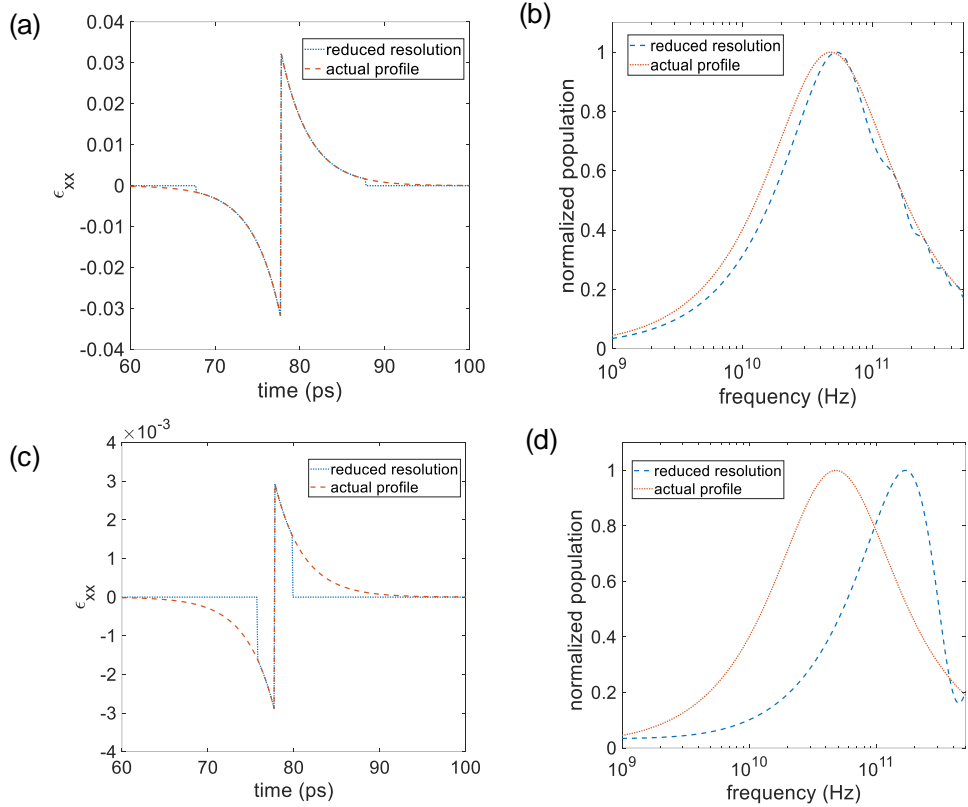

**Fig 4.** Implications of reciprocal space resolution in DFXM imaging of strain wave in diamond produced by ultrafast excitation in aluminum. (a) Presents the time domain profile with infinitesimal reciprocal space resolution (orange dashed line) and with a reciprocal space resolution threshold of 4 x $10^{-4}$ radians, corresponding to a strain resolution of ~1.6 x $10^{-3}$ (blue dotted line) produced using a 22 µJ per pulse laser with a laser spot size of 180 µm in aluminum. (b) Shows the frequency domain reconstruction of the time domain profiles shown in (a). The reconstruction with limited reciprocal space resolution (blue dashed line) is shifted to a slightly higher frequency peak from the actual profile (orange solid line) as low frequency components, which are also lower in amplitude, are not sampled with limited reciprocal space resolution. (c) shows the time domain profile with infinitesimal reciprocal space resolution (orange dashed line) and with a reciprocal space resolution threshold of 4 x $10^{-4}$ radians, corresponding to a strain resolution of ~1.6 x $10^{-3}$ (blue dotted line) produced using a 2.2 µJ per pulse laser with a laser spot size of 180 µm in aluminum. (d) shows the frequency domain reconstruction of the time domain profiles shown in (a). In the case of lower per pulse energy, the frequency domain reconstruction is significantly shifted as only the highest frequency components, which are also high in magnitude, are sampled.

*Finite X-ray beam width*: Fig. 5a shows the concept of a pixel gauge volume (the area inside the red trapezoid in the figure) in DFXM when an illumination beam has a finite thickness. As shown in Fig. 5a, a single pixel in the DFXM image samples the entire projection of the incident beam along the imaging direction into the same pixel (the region shown by the red outline). Therefore, a finite beam width along the direction of the phonon propagation defines an integration volume that defines the minimum resolvable phonon frequencies similarly to the effect of the finite spatial resolution. In the lab coordinate system, the projection of the beam width ($\Delta z_\ell$) into a single pixel of the detector is $\Delta z_\ell \tan 2\theta$. The effective spatial resolution in the object plane is different compared to the resolution when using an infinitesimally thin beam and is specific to the geometry of the experiment. In the geometry used in our simulation, the effective spatial resolution in the grain coordinate system is given by $\Delta x_g \sim \sqrt{\left(\frac{x_p}{2\cos\theta}\right)^2 + \left(\frac{\Delta z_\ell}{\cos\theta}\right)^2}$. Therefore, even when using a pixel size small enough to not significantly affect the frequency reconstruction, the use of a wide pencil beam affects the spatial and frequency resolution significantly in the geometry of our simulation.

In Fig. 5 (b-c) we illustrate the shift in resolvable phonon frequencies for DFXM imaging using a 500-nm thick beam as compared to an infinitesimally thin beam, assuming an effective pixel size (i.e. $\frac{pixel\ size}{magnification}$) of 50 nm. In the specific geometry of the experiment, a 500 nm beam enlarges the gauge volume (integrated diffraction signal from the sample volume contributing to each pixel) of the sampling, resulting in a spatial resolution along $x_g$ to $\sim \frac{500}{\cos\theta_0}$ nm $\sim$ 480 nm. The effective profile reconstructed from DFXM is differs from the actual strain profile in both time and frequency and is effectively modeled assuming an average strain with a spatial resolution of 500 nm.

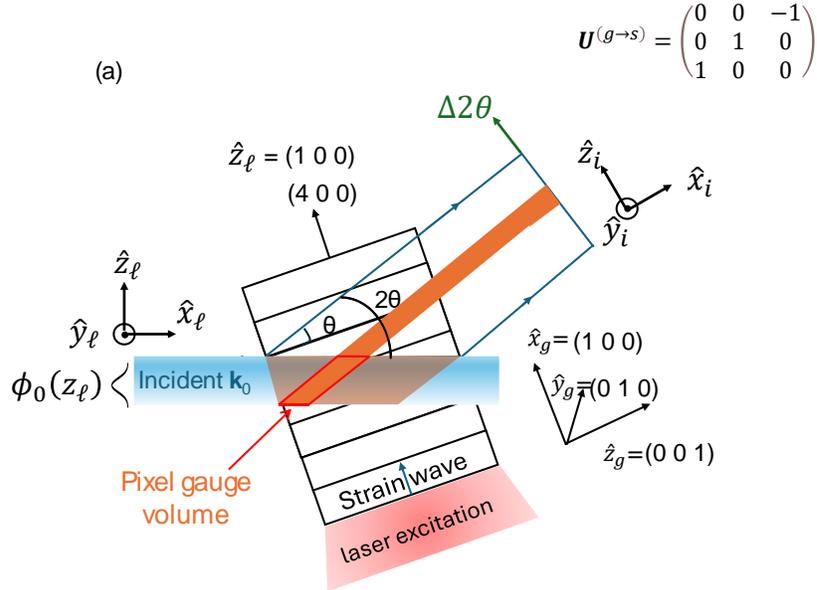

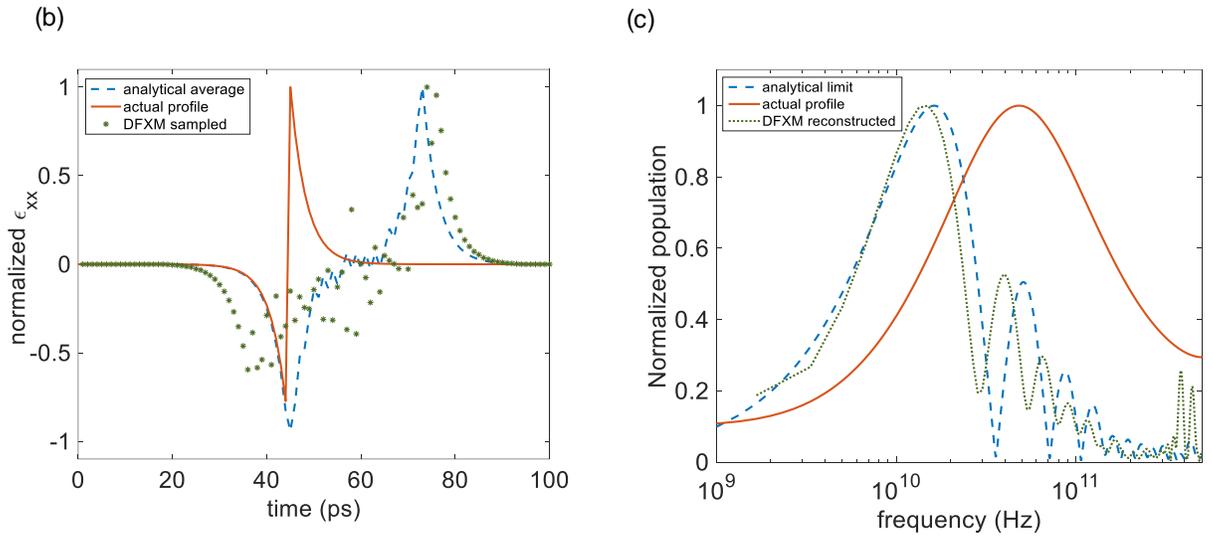

**Fig 5.** Comparison of strain profile in diamond produced with ultrafast excitation in aluminum. (a) Schematic illustrating how the increase in effective pixel size in DFXM is altered by the width of the illuminating line beam. The time-domain plots quantifying this effect is shown in (b) for the time domain of the profile using Gusev-Wright model comparing an assumed infinite spatial resolution (orange solid line) to the strain sampled by DFXM for a 500-nm thick line beam and 50-nm effective pixel size. Blue dotted lines show the analytical average of the strain field with a 500-nm spatial resolution. The corresponding frequency resolution is plotted in (c) comparing the actual profile (orange solid line) to the DFXM reconstructed profile (green dotted line) and the analytical profile with a 500-nm spatial resolution (blue dashed line).

The plots shown in Fig. 5 quantify the aliasing effects described above that set the upper threshold for phonon frequencies that can be measured by DFXM. The plots in Fig. 5c illustrate that the reconstruction of higher frequency phonons requires a thin incident beam width that must be at least $\frac{1}{2\nu}$ the size of the highest frequency that is of interest to the science being studied. While this may seem to require a practically infinitesimal beam width, we note that the narrow beam at the sample compromises the vertical divergence of incident beam that may ultimately reduce the reciprocal-space resolution. As such, the smaller beam width offers the ability to resolve higher frequency phonons at the cost of the ability to resolve low frequency ones. The optimal beam width depends on the expected frequency spectrum – for transducers peaking at lower phonon frequencies (e.g. gold), wider incoming beam with less divergence is better, while for transducers with higher phonon frequencies, small incoming beam width needs to be emphasized. We note that this analysis emphasizes the importance of designing the optical system for DFXM in tandem with the sample topology to ensure that the phonon frequencies of interest are both achievable and resolvable with the experimental setup.

*Finite sampling of the rocking curve*: If a rocking curve scan is collected with sampling steps that are larger than the reciprocal-space resolution of the microscope, the effective reciprocal-space resolution will ultimately be the step size of the rocking scan. Ideally, the step size in a rocking scan must be as close to the instrumental reciprocal space resolution as possible. Furthermore, the sampling in the rocking direction also must cover a sufficiently large range to sample the high amplitude, high frequency components of the strain fields present in the sample – making the small step size and large number of steps difficult for the limited time available at XFEL experiments. The optimal sampling strategy must therefore be selected based on the transducer thickness and material to ensure that the reconstruction afforded by the real- and reciprocal-space resolutions are commensurate with the phonon physics of the material.

### 3.4 Opportunities on imaging the dynamics of GHz acoustic phonons with DFXM

Our simulations have enabled us to understand the tradeoffs in real and reciprocal space resolution in frequency resolved DFXM imaging of acoustic phonons produced by ultrafast excitation of metal transducers. In general, when designing experiments with transducers whose phonon spectrum peaks at high frequencies (and high strain amplitude in time domain), a better real space resolution with a thin sheet beam should be emphasized with the compromise in reciprocal space resolution. On the other hand, for experiments involving transducers with phonon spectrum peaking at low frequency, the requirements on beam width and real space resolution are relaxed, and the emphasis should be to obtain optimal reciprocal space resolution.

Our simulations show that the dynamics of coherent acoustic phonons up to 30 GHz produced by ultrafast excitation of metal transducers can be studied with the current capabilities of DFXM at XFELs. As such, for phonons up to 30 GHz frequencies, DFXM provides a unique opportunity to reconstruct the entire spectrum of populated phonon frequencies with real-space images that can quantify the decay of each frequency as the wavepacket propagates. Previous measurements, primarily using picosecond ultrasonics, have been performed at single frequency values owing to signal-to-noise ratio in the measurements. Beyond the signal to noise considerations, picosecond acoustic measurements require measurement of the acoustic pulses at the surface, which requiring experimental controls like identical laser parameters and surface preparation for multiple sample thickness and delay times. The potential to measure decay constant of coherent acoustic phonon across the phonon entire spectrum with a single measurement promises to add value in the fields of telecommunication and high frequency ultrasound imaging. Till present day, calculations of coherent phonon damping constant beyond specific frequencies have relied on crude estimates. In addition, the ability to simultaneously visualize coherent phonon propagation as well as the strain fields resulting from dislocation could enable to isolate and study phonon dislocation interaction, which would not be possible with any surface measurements.

### 4. Spatially resolved diffuse scattering imaging with DFXM

The coherent phonon modes, described in Section 3, result in a non-zero value of the macroscopic lattice displacement, i.e., $<x> \neq 0$ (Lindenberg, 2001). In the case of coherent acoustic phonons, the result is a shift in Bragg peak that can be measured as strain with DFXM. By contrast, an incoherent distribution of phonons results in zero macroscopic lattice displacement, i.e. $<x> = 0$, meaning that incoherent acoustic phonons do not shift the position of the Bragg peak. Incoherent phonons instead result in a distribution of diffraction intensity around the main Bragg peak, as any non-zero net square lattice displacement, $<x^2> \neq 0$, shifts some diffracted intensity away from the main Bragg peak into the diffuse continuum (Xu and Chiang, 2005). At any specific point in reciprocal space, the scattering vector $\vec{Q}_{sc}$ of the diffuse scattering is related to a corresponding scattering phonon wavevector $\vec{Q}_{ph}$ by $\vec{Q}_{sc} = \vec{Q}_{ph} + \vec{G}$, where $\vec{G}$ is the nearest reciprocal lattice vector. Therefore, the measurement of diffuse scattering intensity at a specific point in reciprocal space away from the Bragg peak enables the measurement of a phonon population of a specific wavevector, even if the population is incoherent.

A significant advantage of DFXM as a lens-based imaging technique is that it offers the possibility to obtain real space resolution to both the Bragg diffracted and diffusely scattered X-rays with excellent $q$-resolution. This is because the lens' aperture acts as a filter in reciprocal space, while the focusing capabilities of the lens also constructs a real-space image when the imaging conditions are satisfied (described in Section 3.1 of Simons et al., 2017). Imaging diffuse scatter could enable DFXM to have applications in imaging the incoherent phonon populations at specific phonon wavevectors that are not near any distinct diffraction peak.

In the following subsections, we describe the adaptation of the formalisms developed in Section 3 to describe the relevant scattering geometry to select diffuse scattering from a specific wavevector and implementation of its imaging with DFXM. We focus on thermal diffuse scattering for its possible importance in applications of thermometry and thermal transport measurements. Our formalism can also be generalized to any phonon population corresponding to a specific phonon wave vector even in non-thermal equilibrium.

### 4.1 Thermal diffuse scattering geometry and phonon structure factor

For X-ray scattering in the lab coordinate system, the X-ray scattering vector $\vec{Q}_{sc}$ is given by $\vec{Q}_{sc} = \vec{G}_\ell$ where $\vec{G}_\ell$ represents the reciprocal lattice vector of the specific plane imaged in the lab coordinate system and is given by (Poulsen et al., 2017) as

$$\vec{G}_\ell = Q_0 \begin{bmatrix} -\sin\theta_0 \\ -\cos\theta_0 \sin\eta_0 \\ \cos\theta_0 \cos\eta_0 \end{bmatrix}, (7)$$

for the nominal scattering angle $\theta_0$ and a nominal azimuthal scattering angle of $\eta_0$. The magnitude of the scattering vector, $Q_0$ is given by $Q_0 = \frac{4\pi}{\lambda}\sin\theta_0$ for an X-ray wavelength $\lambda$.

For a phonon of a specific wave vector $\vec{Q}_{ph}$ defined in lab coordinate system, the scattering vector for diffuse scattering is given by (Poulsen et al., 2017; Xu & Chiang, 2005)

$$\vec{Q}_{sc} = \vec{Q}_{ph} + \vec{G}_\ell = \vec{Q}_{ph} + Q_0 \begin{bmatrix} -\sin\theta_0 \\ -\cos\theta_0 \sin\eta_0 \\ \cos\theta_0 \cos\eta_0 \end{bmatrix}. (8)$$

As we define the scattering angles for nominal Bragg scattering $\theta_0$ and $\eta_0$ in the lab-coordinate system, we find it convenient to describe the scattering geometry for diffuse scattering using the lab coordinate system.

In the lab coordinate system, a general scattering vector $\vec{Q}_{sc,\ell}$ is defined by general scattering angles $\theta$ and $\eta$ by (Poulsen et al., 2017)

$$\vec{Q}_{sc,\ell} = \frac{4\pi}{\lambda}\sin\theta \begin{bmatrix} -\sin\theta \\ -\cos\theta \sin\eta \\ \cos\theta \cos\eta \end{bmatrix}. (9)$$

This equation takes into account the offsets in $\theta$ and $\eta$ resulting from a diffuse scattering specific to a phonon's wavevector. Using Equations (8) and (9), one may select a specific phonon wavevector in the lab coordinate system and calculate the general scattering angles $\theta$ and $\eta$ at which the diffuse scattering intensity will appear.

In the kinematic approximation, the diffuse scattering intensity at a particular scattering vector is given by (Trigo et al., 2010)

$$I(\vec{Q}_{sc,\ell}) \propto \sum_j \frac{1}{\omega_j} \left[n_j + \frac{1}{2}\right] \left|F_j(\vec{Q}_{sc,\ell})\right|^2, (10)$$

where the summation over $j$ describes the sum over the different phonon branches, and $\omega_j$ and $n_j$ represent the phonon frequency and population, respectively, of the $j^{th}$ phonon branch at the wavevector $\vec{Q}_{ph}$. The term $F_j(\vec{Q}_{sc,\ell})$ is the phonon structure factor of the $j^{th}$ phonon branch and is given by

$$F_j(\vec{Q}_{sc,\ell}) = \sum_m f_m e^{-M_m} \left(\vec{Q}_{sc,\ell} \cdot \vec{e}_{s,j,\vec{Q}_{ph}}\right) e^{i(\vec{G}_g \cdot \vec{r}_{m,g})} (11).$$

where the sum over $m$ accounts for contributions from all the basis atoms in the unit cell, $M_m$ is the Debye-Waller factor, and $\vec{r}_{m,g}$ the position of the $m^{th}$ atom in the crystal basis and $\vec{G}_g$ is the reciprocal lattice vector in the grain coordinate system. We further define $\vec{e}_{s,j,\vec{Q}_{ph}}$ as the phonon polarization vector of the $m^{th}$ basis of the $j^{th}$ phonon branch.

In general, X-ray diffuse scattering intensity at a specific scattering vector $\vec{Q}_{sc}$ includes contributions from phonons that have multiple polarizations and arise from multiple branches of the dispersion relations. One can, however, along a high symmetry direction, one can select specific wavevectors in reciprocal space for which the dot product $\vec{Q}_{sc,\ell} \cdot \vec{e}_{s,j,\vec{Q}_{ph}}$ approaches 0 for all except a single phonon polarization, allowing the imaging of that polarization uniquely. For example, when $\vec{Q}_{sc,\ell}$ is parallel to $\vec{G}_\ell$, the resulting dot product $\vec{Q}_{sc,\ell} \cdot \vec{e}_{s,j,\vec{Q}_{ph}} = 0$ for the transverse polarizations, meaning that signal measured at $\vec{Q}_{sc}$ describes only the longitudinal phonons in this specific scattering geometry.

**4.2 Implementation of Diffuse Scatter DFXM in thermometry**

The ability to image thermal diffuse scattering from a specific phonon wavevector can be generally applied to phonons in thermal equilibrium or under non-equilibrium conditions. In this work, we present one specific example of imaging phonon populations in local thermal equilibrium for thermometry in simplified DFXM geometry to demonstrate the implementation of thermal diffuse scattering imaging into the DFXM simulations.

For thermal phonons, the population $n_j$ of a phonon with frequency $\omega$ at temperature $T$ is given by (Xu and Chiang, 2005)

$$n_j = \coth\left(\frac{\hbar\omega}{2k_B T}\right), (12)$$

where $k_B$ is the Boltzmann constant. For thermometry, one can select a specific wavevector parallel to $\vec{G}_\ell$ such that the contribution of phonons to the DFXM signal arises only from phonons of longitudinal polarization. Our selection of longitudinal phonons in this case, from the acoustic branch, is relevant to thermometry because the contribution from the optical branch has much higher phonon frequencies with significantly smaller populations than the same frequency acoustic modes. As thermometry is most sensitive at a phonon frequency $\omega \approx \frac{k_B T}{\hbar}$, our implementation in this section focuses on a phonon frequency of $f = 6$ THz ($\omega$ =37.68 rad/s), corresponding to $T \approx 300\ K$.

For a 6 THz longitudinal acoustic phonon in diamond parallel to (1 0 0) direction and imaged near a (4 0 0) peak in the Simplified Geometry, we use Equations (7) and (8) to determine that $\Delta\theta = 0.027$ radians and $\Delta\eta = 0$ radians from the nominal Bragg condition. Since 0.027 radians is larger than a typical CRL aperture for the imaging distances assumed in this work (typically ~$10^{-3}$ radians), one would need to translate the CRL and detector positions by 0.027 radians about $2\theta$ for DFXM to image the selected phonon wavevector. Lens and detector translations like the one defined here are generally required to image diffuse scattering for high-frequency phonons for which the difference in the scattering angle is greater than the aperture of the CRL. The detector position must change as well as the CRL to ensure effective alignment and prevent imaging artifacts like warpage and distortion that arise from misaligned imaging lenses.

With the model described in Section 4.1, the formalism to image thermal diffuse scattering from a specific wavevector, is only different from the usual DFXM formalism by accounting for the scattering probability determined by the phonon population and the Debye-Waller factor. In the simulation we perform for thermometry shown in Fig. 6, we introduce a scattering probability in each real space grid in our simulation scaling with the phonon population described by Equation 12, i.e.

$$Probability_T(x_g, y_g, z_g) = \coth\left(\frac{\hbar\omega}{2k_B T(z_g)}\right) \quad (13).$$

We ignore the Debye-Waller factor in this scaling as it changes by less than 1% for the range of temperature and the peak used in the simulation. Fig. 6 shows a DFXM simulation of the 6 THz acoustic phonon in diamond, using the simplified geometry. For these simulations, we use the identical reciprocal space resolution function described in Section 3, where the identity matrix is used to describe the deformation gradient tensor, $\boldsymbol{F}^g$. This formalism can be generalized to non-thermal phonons replacing the phonon population determined by temperature to any phonon population corresponding to a specific wavevector.

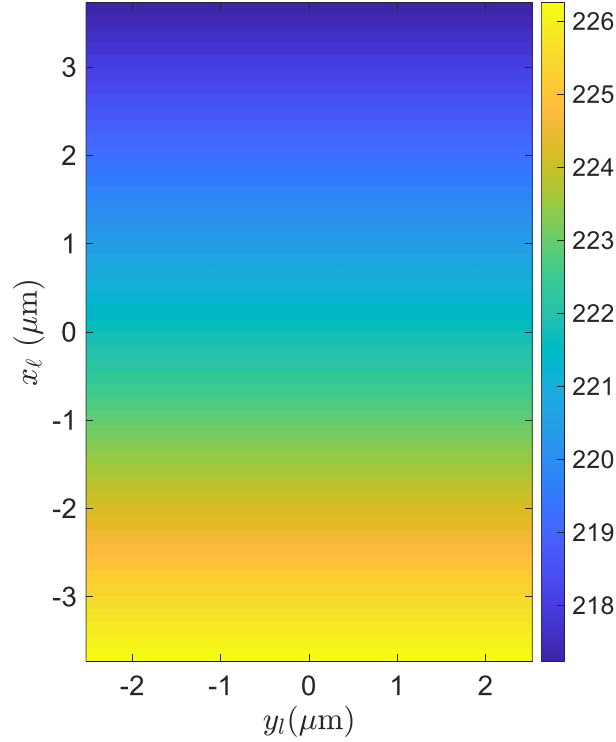

**Fig 6.** Thermal diffuse scattering intensity due to 6 THz thermal phonons simulated DFXM in diamond, assuming a linear temperature gradient of 10 K/μm along the $\hat{z}_g$ axis, corresponding to a temperature gradient of ~2.5 K/ μm along the $\hat{x}_l\ axis$ . The colorbar intensity units are arbitrary units corresponding to diffuse scattering intensity, as measured by DFXM intensity.

The benefits of these simulations, and the implications of real and reciprocal space resolution and finite beam width are discussed in the following subsections.

### 4.3 Real and Reciprocal space resolutions in imaging diffuse scattering with DFXM

*Real space resolution in imaging:* For any application of imaging diffuse scattering, the achievable resolution in real space is almost identical to that obtained from a Bragg scattering experiment with DFXM. The finite value of the beam width $\Delta z$ has the same implication to real space resolution as in the case of coherent phonons i.e. the effective resolution in real space in the lab coordinate is ~ $\sqrt{x_p^2 + \Delta z^2 \tan 2\theta}$ . At synchrotrons, the achievable spatial resolution is ~150 nm while at XFELs, this is ~ 300 nm.

*Reciprocal space resolution*: Unlike the case of Bragg scattering, where both the crystal and the objective lens act as filters in the reciprocal space, the reciprocal space resolution of diffuse scattering is entirely determined by the aperture of the objective (Poulsen et al., 2017). Typically, the numerical aperture of the CRL ~ $10^{-3}$ radians. Therefore, the reciprocal space resolution in imaging diffuse scattering is limited to $10^{-3}$ radians, though the resolution along one axis may be enhanced by the beam divergence because of the resolution function, as described in Holstad, et al. (2020).

The finite reciprocal-space resolution also means that the simulation presented in Fig. 6 is not entirely accurate. In general, one must calculate the intensities in reciprocal space around the entire lens aperture, not simply the contribution at the center as calculated in Fig. 6 for only 6 THz. Such calculation requires a computation of the contribution from every neighboring wavevector with every polarization, which is beyond the scope of this work. We note, however, that our simulation provides a reasonable estimate of the DFXM image produced by thermal diffuse scatter, as the phonon dispersion does not significantly change over the largest $10^{-3}$ radian range considered in the DFXM forward model. For more applications requiring more precise sampling of wavevectors, however, the estimated contribution from the entire reciprocal space spanned by the CRL must be calculated.

The theoretical limit to the real-space resolution is given by $\Delta x \approx \frac{\lambda}{3\sigma_a}$, while the reciprocal-space resolution scales linearly with the numerical aperture $\sigma_a$. Therefore, especially for the case of diffuse scattering imaging, we emphasize the opportunity for optics development allowing imaging at high X-ray energies which would enable the improvement in spatial resolution without compromising reciprocal space resolution is possible.

### 4.4 Opportunities offered by spatially resolved diffuse scattering imaging with DFXM

In terms of thermometry alone, the ~150-nm spatial resolution achievable with DFXM would only offer an order of magnitude improvement over what is currently possible through the depth of a sample with X-ray topography or a rastered X-ray diffraction scan (Chalise et. al, 2022). In the field of thermal transport measurement, achieving spatially and temporally resolved imaging of phonon population at nanometers length scale through the depth of the sample provides opportunities for fundamental advancement of the knowledge of heat transfer not possible with currently existing surface measurements (e.g. Time domain thermoreflectance (Cahill, 2004)) which typically assume Fourier's law of heat diffusion. Imaging diffuse scattering with ~100 nm spatial resolution in time resolved measurements would offer significant improvements to existing measurements of diffuse scattering, which are resolved only in reciprocal space (Trigo et al. 2013), and are limited in providing the information on spatial evolution.

## 5. Discussion

DFXM offers the possibility to obtain real-space images of coherent and incoherent phonons through the depth of the sample with resolution in real space on the order of ~150 nm and in reciprocal space of ~$10^{-3}$ radians (thermal diffuse scattering) to $10^{-4}$ radians (coherent phonons). The combination of real-space imaging while filtering the scattered light in reciprocal space makes DFXM unique as compared to X-ray techniques like X-ray topography (Danilewsky, 2020) and classical diffuse scatter (Lindenberg et al., 2000). Finally, DFXM's ability to image dynamics through the thickness of macroscopic crystals is not possible in phonon measurement techniques like picosecond ultrasonics, offering crucial insights into the processes of Akheizer damping, decoherence and phonon-dislocation interactions which are important in the fields of communication, quantum information and ultrasonic imaging.

From the application of the classical photothermal models in Section 3.1 to the DFXM forward model introduced in Section 3.2, we demonstrate the utility of DFXM in selectively imaging specific frequency components of the coherent phonons generated by ultrafast photothermal excitation in a transducer material. We illustrate that in diamond, frequency resolved measurements up to 30 GHz is possible with existing DFXM capabilities, potentially advancing our knowledge of phonon-dislocation interaction and phonon-damping. Our analysis also demonstrates the importance of maintaining high spatial resolution to be able to accurately resolve the high-frequency phonon components. The balance between the need for high spatial resolution and high reciprocal-space resolution to accurately resolve the low-frequency phonon components highlight the need to select the optics for an experiment based on the phonon frequencies most important to each application. Our findings also demonstrate an opportunity for future developments in DFXM instrumentation to explore focusing optics (e.g. CRL, Laue lenses, etc.) for high photon energies that can still offer the high numerical apertures and magnifications needed to reconcile the present frequency tradeoffs of measuring coherent phonons.

Imaging diffuse scattering with DFXM opens a range of possibilities which have not been fully discussed in the paper. 3D thermometry with sub-micron resolution, which could be one of the applications of diffuse scattering imaging, itself has significant technological applications (Chalise et al., 2022 and Chalise & Cahill, 2023). Further, thermal transport measurements, which could be possible with the implementation of DFXM with time resolved x-ray diffraction could provide information not possible with surface based thermal metrologies using optical (Cahill, 2004 and Zheng et al., 2022) or electrical (Chalise et al., 2023) heating and readout. Additionally, non-thermal phonon populations can also be imaged with diffuse scattering imaging (Trigo et al., 2013).

We note that our present work introduces material models to describe phonons for DFXM but leaves open several key items that we envision will follow this work. For example, the photometric efficiency of DFXM signals generated by X-ray scattering produced by phonons away from the main Bragg peak must be considered to ensure adequate signal-to-noise for experiments. While this is crucial to consider, the scattering intensity is highly system dependent and thus is beyond the scope of the present formalism-focused work. We also note that the results presented in this work are purely derived using kinematic diffraction theory for both coherent and incoherent phonons. Results by Lindenberg et al. (2000) have shown that dynamical diffraction effects in scattering of X-rays by coherent phonons yield results not explained by the kinematic theory. Dynamical diffraction effects have also been demonstrated to be of higher relevance to DFXM experiments at XFELs due to the high coherence of the X-ray beams (Irvine et al., 2024). Our future work will address DFXM imaging of coherent phonons through the consideration of the dynamical theory of diffraction.

## 6. Conclusion

In this work, we review the relevant models to define the strain profiles for GHz acoustic phonons and define their contributions to imaging signal in DFXM. We demonstrate the application of this model using the DFXM formalism developed using kinematic diffraction theory and geometrical optics (Poulsen et al., 2023) and use this to describe quantitative frequency resolved measurements of coherent acoustic phonons. Our work demonstrates the application of this approach to understand discrete phonon scattering and damping relevant to acoustic filters and high frequency ultrasound imaging. We then expand this approach to phonon population measurements based on DFXM for thermal diffuse scatter to enable applications in imaging thermometry and non-equilibrium thermal transport. In both cases, we discuss the tradeoffs between real- and reciprocal-space resolutions and discuss the experimental opportunities enabled by these measurements.

## Acknowledgements

The authors acknowledge Prof. Aaron Lindenberg (Stanford University) for valuable discussions on thermal diffuse scattering, and Dr. Alexei Maznev (MIT), Prof. Keith Nelson (MIT), and Prof. David Cahill

(UIUC) for discussions on imaging coherent phonons. This study was supported by the Department of Energy, Office of Science, Basic Energy Sciences, Materials Sciences and Engineering Division, under Contract DEAC02-76SF00515.